\documentclass[12pt]{article}

\newcommand{\x}{arXiv:}
\newcommand{\m}{\mathrm}

\usepackage{graphicx}

\begin{document}
\thispagestyle{empty}
\begin{center}

\null \vskip-1truecm \vskip2truecm

{\Large{\bf \textsf{Kerr Black Holes Are Not Fragile}}}

{\large{\bf \textsf{}}}

\vskip0.5truecm

{\large \textsf{Brett McInnes
}}

\vskip0.5truecm

\textsf{\\ Centro de Estudios Cient}$\acute{\textsf{i}}$\textsf{ficos (CECs), Valdivia, Chile}
\vskip0.1truecm
\textsf{\\ and}
\vskip0.1truecm
\textsf{\\ National
  University of Singapore}\footnote{Permanent Address}
  \vskip0.3truecm
\textsf{email: matmcinn@nus.edu.sg}\\

\end{center}
\vskip1truecm \centerline{\textsf{ABSTRACT}} \baselineskip=15pt

\medskip
Certain AdS black holes are ``fragile", in the sense that, if they are deformed excessively, they become unstable to a fundamental non-perturbative stringy effect analogous to Schwinger pair-production [of branes]. Near-extremal topologically spherical AdS-Kerr black holes, which are natural candidates for string-theoretic models of the very rapidly rotating black holes that have actually been observed to exist, do represent a very drastic deformation of the AdS-Schwarzschild geometry. One therefore has strong reason to fear that these objects might be ``fragile", which in turn could mean that asymptotically flat rapidly rotating black holes might be fragile in string theory. Here we show that this does not happen: despite the severe deformation implied by near-extremal angular momenta, brane pair production around topologically spherical AdS-Kerr-Newman black holes is always suppressed.

\newpage

\addtocounter{section}{1}
\section* {\large{\textsf{1. The Consequences of Distorting AdS}}}
The geodesics of the Kerr spacetime \cite{kn:visser} are such that there is an innermost stable circular orbit, the existence of which has a strong effect on the accretion disc of a spinning black hole. Using this fact, observers \cite{kn:McC} have concluded that several known black holes, such as the one belonging to the binary system GRS 1915+105, may have an angular momentum per squared mass very close to the maximal value permitted by cosmic censorship. We are therefore in a potentially very interesting position: we have \emph{an actual observed object} which, granted cosmic censorship, defines a spacetime that is almost maximally different from Minkowski spacetime. It is quite possible that observations of this object may eventually demand a description in terms of physics beyond the classical level, perhaps using the Kerr/CFT duality \cite{kn:guica}\cite{kn:andystrom} in the form it takes near to but away from extremality \cite{kn:bredberg}\cite{kn:carlip}.

One naturally wonders whether specifically string-theoretic effects might be relevant under these extreme conditions. In particular, one would like to be reassured that non-perturbative string-theoretic effects do not \emph{destabilize} rapidly rotating black holes, given that we now know that such objects do exist. The natural arena to investigate such questions \cite{kn:krishnan} is the spacetime of a large, near-extremal AdS-Kerr black hole\footnote{``Large" here refers to the familiar fact that AdS black holes have good thermal behaviour when the event horizon is not too small relative to the asymptotic curvature radius L. Large AdS black holes also have the virtue of being immune to superradiance: see for example \cite{kn:dias}. Ultimately, however, one is also interested \cite{kn:david} in ``small" AdS black holes, so we shall not neglect those.}: the hope is that the lessons learned by studying string theory on such a background might have implications for realistic black holes. This is, for example, the reasoning that underlies the standard string-theoretic view of the black hole unitarity paradox \cite{kn:juan}\cite{kn:vijay}.

The primary tool for using string theory in this manner is of course the AdS/CFT correspondence. As has been emphasised from the beginning \cite{kn:confined}, the correspondence applies to \emph{any} asymptotically AdS spacetime, so it can be used to study the consequences of moving away from pure AdS geometry in the bulk. AdS itself, and also AdS-Schwarzschild, are expected to be ``well-behaved", in the sense that they have well-behaved AdS/CFT duals. By the same token, however, the AdS/CFT correspondence suggests ---$\,$ indeed, in some cases \emph{demands} ---$\,$ that, if we distort these spacetimes in certain ways, then they must become unstable in some manner that is not obvious from the bulk point of view and which must have a purely ``stringy" origin. [``Distortion" here will be taken to include variations of the ADM parameters, since such variations affect the spacetime geometry indirectly; it can also cover direct modifications of the geometry or topology.]

For example, suppose that we distort the AdS-Schwarzschild black hole in the following manner. First, we change the topology of the event horizon from that of a sphere to that of a flat space, [say] a flat torus \cite{kn:lemmo}; then, we gradually increase the electric charge on the black hole. These AdS-Reissner-Nordstr$\ddot{\m{o}}$m black holes with flat event horizons are thought to be dual \cite{kn:chamblin} to a field-theoretic configuration which, for some purposes, can be used to model a quark-gluon plasma \cite{kn:edshur} inhabiting a locally flat spacetime. Now such a plasma cannot exist at arbitrarily low temperatures; as its temperature is lowered, it passes through either a crossover or a phase transition to some other state [confined at low values of the chemical potential, deconfined at higher values]. The AdS/CFT correspondence therefore requires that AdS black holes with flat event horizons must necessarily suffer from some kind of instability at some point as their charge is increased [since that lowers the temperature], so that they never reach extremality [which corresponds to zero temperature]. But no such instability appears in ordinary gravity; pure AdS-Reissner-Nordstr$\ddot{\m{o}}$m black holes seem to be well-behaved for all values of the charge up to the extremal value. Therefore: \emph{there must be some ``stringy" effect which causes these black holes to become unstable when they are distorted excessively in this manner.}

To take another example: one can distort the ordinary, uncharged AdS-Schwarzschild geometry [with a spherical event horizon] in a more direct way, by distorting the sphere at infinity that corresponds to the event horizon, and then solving the field equations. This was done in the work of Murata et al. \cite{kn:murata} [see also \cite{kn:kunz}]. One of the more remarkable findings of \cite{kn:murata} was the following: if one distorts the event horizon of the black hole so that it becomes a prolate spheroid, then beyond a certain point the scalar sector of the boundary theory becomes unphysical [the energy ceases to be real for some parameter choices]. The authors point out that this is puzzling, because, as in the previous example, there seems to be no reason to think that the black hole should cease to be a physical solution under this [as it turns out, rather mild] deformation. Again, the conclusion is that some stringy effect must be sensitive to these geometric distortions of the black hole. [We note in passing that this effect, whatever it may be, is a critical component of the AdS/CFT correspondence; \emph{without it, the duality would fail}.]

The moral to be drawn from these examples is straightforward: string-theoretic corrections imply that at least some AdS black holes are ``fragile". But this raises an immediate, very basic question with regard to the idea that AdS black holes might be used to study objects like the black hole in GRS 1915+105: do near-extremal AdS-Kerr-Newman black holes even \emph{exist} as stable objects in string theory? To see that there are serious grounds for concern here, note that Kerr-Newman black holes can differ very drastically from Schwarzschild black holes; they differ \emph{both} in the radial direction [as in the first example above] \emph{and} in the angular directions [as in the second example].

For example, the intrinsic metric on the event horizon of an AdS-Kerr[-Newman] black hole has the form
\begin{equation}\label{ALPHA}
\m{g(AdSKN)_{eh} =  {r_{eh}^2\;+\;a^2cos^2\theta \over 1 - {a^2\over L^2} \, cos^2\theta}d\theta^2 \;+\;{sin^2\theta \,(1 - {a^2\over L^2} \, cos^2\theta) \over r_{eh}^2\;+\;a^2cos^2\theta}\Bigg[{r_{eh}^2\,+\,a^2 \over 1 - {a^2\over L^2}}\Bigg]^2 d\phi^2},
\end{equation}
where r$_{\m{eh}}$ is the value of a radial coordinate at the event horizon, a is related to the angular momentum, and L is the asymptotic AdS curvature radius. Even for moderate values of a, this is clearly a severe distortion of the metric of the event horizon of an AdS-Schwarzschild black hole.
For example, the Gaussian curvature of g(AdSK)$_{\m{eh}}$ is actually \emph{negative} for sufficiently near-extremal holes over certain regions of the event horizon, a clear indication of a strong deformation of the round two-sphere. [This is discussed in detail in the asymptotically flat case ---$\,$ where ``near-extremal" for this purpose means that the angular momentum per squared mass exceeds $\approx$ 0.866 ---$\,$ in \cite{kn:visser}; a similar statement is valid in the asymptotically AdS case also, though the details differ\footnote{We are not claiming that the Gaussian curvature of the event horizon has any particular significance: we merely wish to emphasise that adding large amounts of angular momentum has a more dramatic effect on the geometry than intuition might suggest. Note that the Gaussian curvature of a topologically spherical \emph{two-dimensional} event horizon cannot be negative everywhere, by the Gauss-Bonnet theorem.}.] Compare this with the deformations considered by Murata et al.: in that case, a mere ``squashing", resulting in an event horizon which is simply a [not very] prolate sphere, gives rise to a serious instability. Since we know [from AdS/CFT duality] that the black hole becomes unstable in that case, \emph{there is every reason to fear that near-extremal AdS-Kerr-Newman black holes are likewise unstable}, for similar reasons.

In order to settle this matter, we must of course identify the stringy effect responsible for enforcing the AdS/CFT duality in the above two examples. We now turn to this.

In string theory, there is a [non-perturbative] brane pair-production instability, first analysed by Seiberg and Witten \cite{kn:seiberg} [see also \cite{kn:wittenyau}], which can arise in asymptotically AdS spacetimes with local geometries sufficiently different from that of AdS itself. Like the familiar Hawking-Page phase transition, this effect is of fundamental importance because, unlike other forms of instability, it does \emph{not} depend on the presence of specific forms of matter [such as scalar fields \cite{kn:yenchin} with potentials specially tailored to trigger it]. Instead, the instability arises as the inevitable consequence of the fact that certain AdS-like geometries induce an instability of objects ---$\,$ branes ---$\,$ which are automatically present in the spectrum of string theory.

In \cite{kn:AdSRN}\cite{kn:triple}\cite{kn:73} we have argued that this fundamental stringy form of instability is the bulk effect required to understand the instability of near-extremal pure AdS-Reissner-Nordstr$\ddot{\m{o}}$m black holes with flat event horizons, as discussed in the first example above; and in \cite{kn:fragile} it was shown that it is the source of the bulk instability corresponding to the boundary instability discussed by Murata et al. \cite{kn:murata}. In these two cases, then, this \emph{Seiberg-Witten instability} is the effect that enforces the black hole instability demanded by the AdS/CFT correspondence.

We can now formulate the above discussion in a more concrete way. Evidently branes, because they are extended objects, are sensitive to the distortions of the AdS-Schwarzschild geometry arising in the examples we have discussed, to the degree that they can render the whole system unstable. The question is whether they are equally sensitive to the apparently severe [but of course quite different] distortion produced by adding angular momentum to the black hole.

Concretely, then, the concern is that the effect of rotation might be to render the action for branes negative over some region around the black hole. If that were to occur, then brane pair-production would ensue, and one would ultimately expect the branes to begin to affect the black hole; presumably the end result would be differential emission of pairs of branes, so that the angular momentum of a near-extremal hole would be reduced until the negative brane action is removed\footnote{In this case, one would still need to study the time scale on which the instability develops; see \cite{kn:barbon}\cite{kn:fragile}.}. \emph{This could be interpreted as meaning that string theory does not admit solutions corresponding to stable, very rapidly rotating black holes}. In view of the fact that the observed black hole in GRS 1915+105 has an angular momentum per squared mass \cite{kn:McC} equal to \emph{at least} 98$\%$ of the extremal value, this would be a disturbing conclusion.

In this work we will show that nothing of this sort happens: topologically spherical AdS-Kerr-Newman black holes are completely stable against the Seiberg-Witten effect, for all angular momenta up to and including extremal values. The unusual geometry near to the event horizon does of course have an effect on the brane action in that region, but, [very] surprisingly, \emph{this effect is never sufficient to give rise to the feared instability}.

We begin with a brief discussion of the AdS-Kerr-Newman spacetime. This is necessary because the definition of extremality in this case takes an unusual form, which we need to understand for our numerical investigations.

\addtocounter{section}{1}
\section* {\large{\textsf{2. Near-Extremal AdS-Kerr-Newman Black Holes}}}
The four-dimensional topologically spherical AdS-Kerr-Newman metric \cite{kn:carter} [see \cite{kn:hawrot}\cite{kn:pope}\cite{kn:emparan} for higher dimensions] is given, in Boyer-Lindquist-like coordinates, by
\begin{equation}\label{A}
\m{g(AdSKN) = - {\Delta_r \over \rho^2}\Bigg[\,dt \; - \; {a \over \Xi}sin^2\theta \,d\phi\Bigg]^2\;+\;{\rho^2 \over \Delta_r}dr^2\;+\;{\rho^2 \over \Delta_{\theta}}d\theta^2 \;+\;{sin^2\theta \,\Delta_{\theta} \over \rho^2}\Bigg[a\,dt \; - \;{r^2\,+\,a^2 \over \Xi}\,d\phi\Bigg]^2},
\end{equation}
where
\begin{eqnarray}\label{eq:B}
\rho^2& = & \m{r^2\;+\;a^2cos^2\theta} \nonumber\\
\m{\Delta_r} & = & \m{(r^2+a^2)\Big(1 + {r^2\over L^2}\Big) - 2Mr + Q^2}\nonumber\\
\Delta_{\theta}& = & \m{1 - {a^2\over L^2} \, cos^2\theta} \nonumber\\
\Xi & = & \m{1 - {a^2\over L^2}.}
\end{eqnarray}
Here $- 1$/L$^2$ is the asymptotic curvature, a is the angular momentum/mass ratio [see below], and M and Q are related to the mass and charge respectively. [We are mainly interested in the case Q = 0; however, non-zero charge does not change our discussion in this section in any important way.]

The physical mass, charge, and angular momentum of the black hole are given, in the notation of \cite{kn:gibperry} [see also \cite{kn:dolan}], by
\begin{equation}\label{BLECH}
\m{E\;=\;M/\Xi^2, \;\;\;\;\;C\;=\;Q/\Xi,\;\;\;\;\; J \;=\; aM/\Xi^2.}
\end{equation}
There are two constraints which must be imposed on these quantities. First, it is clear from the structure of the metric tensor that $\Delta_{\theta}$ must not be allowed to vanish, which means that $\Xi$ must be positive. Hence
\begin{equation}\label{BB}
\m{a^2/L^2 \;<\; 1;}
\end{equation}
this will be important in our subsequent discussion of the brane action.

Notice that E differs very greatly from M when a/L is close to unity, and so M must be interpreted with care; however, the key parameter a may still be interpreted as the angular momentum per unit mass, as in the asymptotically flat case. Thus the inequality (\ref{BB}) means that the asymptotic curvature constrains the angular momentum per unit mass of the black hole.

A more familiar constraint originates in cosmic censorship\footnote{It has been suggested \cite{kn:gimon} that string theory might allow violations of classical censorship, but subsequent investigations \cite{kn:pani} show that such solutions, if indeed they exist, are not likely to be physically relevant.}. In the asymptotically flat case, the dimensionless quantity J/E$^2$ [which might be called the specific angular momentum], conventionally denoted in the astrophysics literature by a$^*$, cannot exceed unity if an event horizon is to exist. This quantity is given here by
\begin{equation}\label{BLECHH}
\m{a^* \;=\;J/E^2 \;=\; {a\over M}\Big(1 - {a^2\over L^2}\Big)^2,}
\end{equation}
a relation which will be useful below.

In the asymptotically flat case, while cosmic censorship requires J/E$^2$ to be no larger than unity, it can take that value [in the extremal case]. Here the situation is more complicated: the value of r at the event horizon, r$_{\m{eh}}$ [the largest root of the polynomial $\Delta_{\m{r}}$], is well-defined only if the various parameters satisfy \cite{kn:caldarelli}
\begin{eqnarray}\label{BBB}
\m{M/L} & \geq &\m{\Gamma \Bigg({a\over L},{Q \over L}\Bigg)\;=\;{1\over 3\sqrt{6}}\Bigg[\sqrt{1+14a^2/L^2+12Q^2/L^2+a^4/L^4}+(2a^2/L^2)+2\Bigg]}
\nonumber \\
& & \;\;\;\;\;\;\;\;\;\;\;\;\;\;\;\;\;\;\;\;\;\times\;\m{\sqrt{\sqrt{1+14a^2/L^2+12Q^2/L^2+a^4/L^4}-(a^2/L^2) - 1}}.
\end{eqnarray}

\begin{figure}[!h]
\centering
\includegraphics[width=0.7\textwidth]{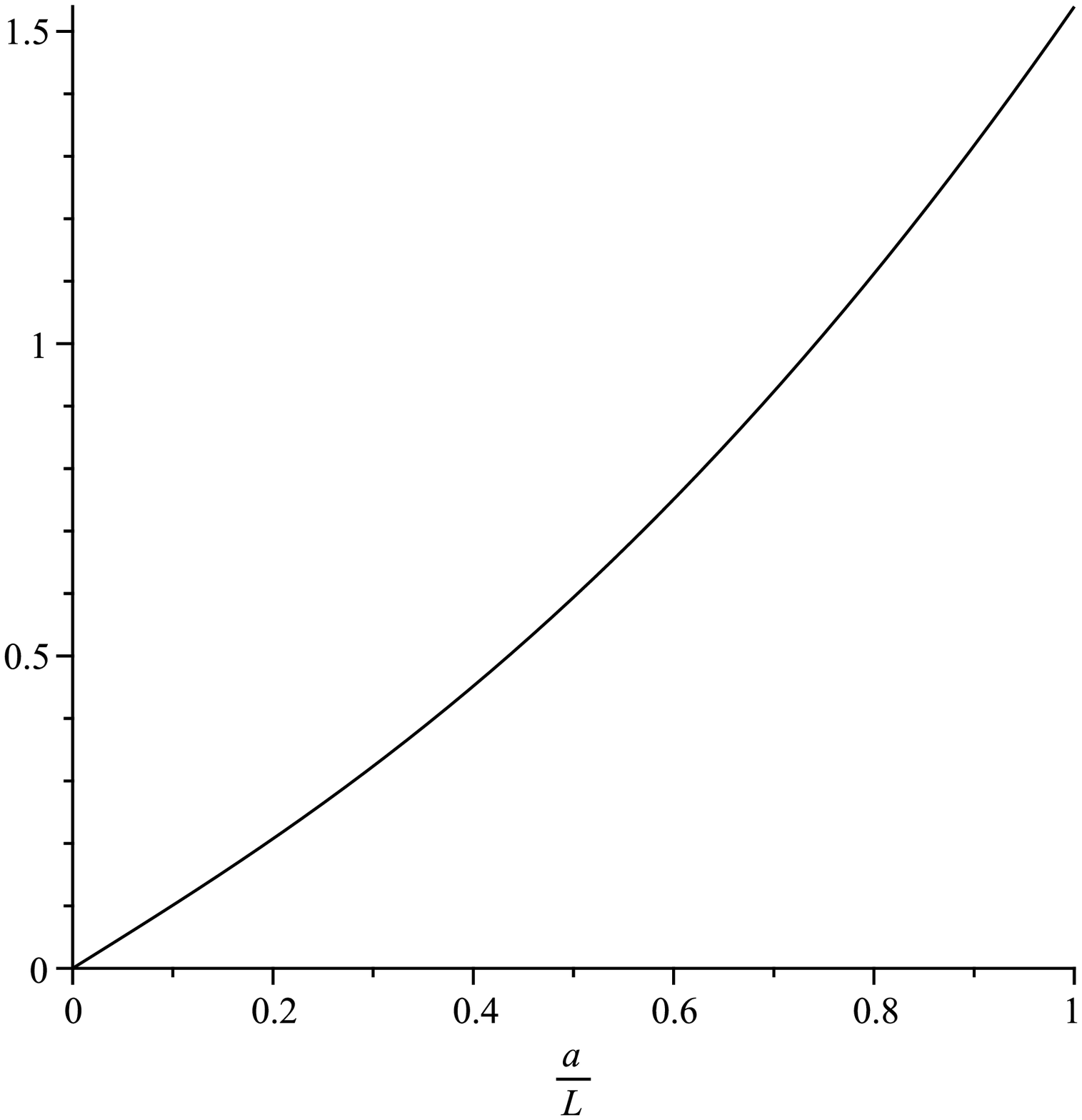}
\caption{Graph of $\m{\Gamma ({a\over L}, 0)}$ }
\end{figure}
A graph of $\m{\Gamma (a/L, 0)}$ is shown in Figure 1. We see that [taking Q = 0] the physical pairs (a/L, M/L) are those which lie above the curve and to the left of the vertical line a/L = 1. As the graph lies entirely above the line with slope unity, it follows that a/M must be strictly less than unity, and then we see from (\ref{BLECHH}) that J/E$^2$ must also be strictly smaller than unity; in fact, it is often substantially smaller even for black holes which are near extremality [in the sense that (\ref{BBB}) is close to being saturated]. That is, cosmic censorship imposes a stricter constraint on the angular momentum [measured relative to the squared mass] in the asymptotically AdS case than in the asymptotically flat case. We will return to this important point below.

It follows from this discussion that we \emph{cannot} usually expect to understand realistic black holes by studying AdS black holes with the same value of J/E$^2$. However, this problem can be solved as follows. What we really want to study are \emph{near-extremal} black holes like the one in GRS 1915+105; it therefore seems reasonable to posit that the relevant AdS objects are AdS-Kerr black holes of the same mass which are ``as close to extremality" as the one in GRS 1915+105 or other similar black holes. Here, ``closeness to extremality" should be interpreted with due regard to Figure 1.

In the specific case \cite{kn:McC} of the black hole in GRS 1915+105, J/E$^2$ is at least 0.98; we can say that the corresponding object in AdS [for a fixed value of L, which is also constrained if one requires that the black hole should be ``large" and that superradiance should be avoided] is an [electrically neutral] black hole with with a$^*$ = J/E$^2$ equal to at least 98$\%$ of its value that one obtains by fixing the physical mass\footnote{Unfortunately, while J/E$^2$ is known rather precisely for this object, the mass itself is not, being constrained only to within about 30$\%$. Other black holes, almost as close to extremality, but with more precisely determined masses, have however been discovered: see \cite{kn:McC2}.} and increasing the angular momentum until (\ref{BBB}) is saturated.

To be more specific: if E is fixed, then increasing J is, since J/E = a, tantamount to increasing a; the effect of that is to decrease M = $\Xi^2$E and to increase $\m{\Gamma (a/L, 0)}$, and the two eventually meet. Suppose therefore that we have an AdS-Kerr metric with given parameters a$_0$ and M$_0$: then the extremal value of a for this black hole, a$_{\m{ext}}$, is found by solving the equation
\begin{equation}\label{BLAHH}
\m{\Bigg({1 - a_{ext}^2/L^2\over 1 - a_0^2/L^2}\Bigg)^2{M_0 \over L} \;=\;\Gamma \Bigg({a_{ext}\over L},0\Bigg).}
\end{equation}
If a$_0^*$ is the value of a$^*$ for this black hole, then
\begin{equation}\label{BLAHHH}
\m{a_0^*/a_{ext}^* \; = \; a_0/a_{ext},}
\end{equation}
and this is the quantity which allows us to judge whether the given black hole is near to extremality.

It is important to realise that a near-extremal AdS black hole may be rotating at a rate which, by asymptotically flat standards, is actually not very large. We will illustrate this with two examples: the first is a ``small" black hole [denoted S], the second is ``large" [denoted L].

{{\bf \textsf{Example S.}}} Take M/L = 1, Q = 0, a/L = 1/$\sqrt{2}$. Then (\ref{BB}) is obviously satisfied, while $\Gamma({1\over \sqrt{2}},0) \approx 0.93612$, so (\ref{BBB}) is also satisfied. The dimensionless parameter J/E$^2$ for this black hole is, from equation (\ref{BLECHH}), approximately  0.17678, which is not particularly large by the standards of observed black holes. But in AdS, this amount of angular momentum means that the hole is nearly extremal: solving (\ref{BLAHH}) with these parameter choices we find a$_{\m{ext}}$ $\approx$ 0.715687 L, and then equation (\ref{BLAHHH}) informs us that a$^*$ is approximately 98.8$\%$ of the extremal value. Thus, even though its specific angular momentum is not very large, the AdS black hole with these parameters is rotating almost as rapidly as cosmic censorship permits. One might say that the addition of angular momentum \emph{has a stronger effect on AdS black hole geometries than on their asymptotically flat counterparts}.

This particular black hole is actually ``small", in the sense that the radius of its event horizon [$\approx$ 0.6377 L] is somewhat smaller than L. While such black holes are not our main concern here, they are not without interest \cite{kn:david}, and one would like to know whether they suffer any additional form of instability beyond the well-understood ones associated with evaporation and superradiance.

{{\bf \textsf{Example L.}}} Take M/L = 2, Q = 0, a/L = 0.98. Again, (\ref{BB}) and (\ref{BBB}) are satisfied, but now the radius of the event horizon is $\approx$ 1.0194 L, so this black hole is ``large". Here a$_{\m{ext}}$ $\approx$ 0.49135 L, and a$^*$ is approximately 99.72$\%$ of the extremal value, so this example is even closer to extremality than the preceding one; yet it is rotating ``slowly" enough to escape being superradiant [see the discussion of ``slow" rotation in \cite{kn:krishnan}].

We see that both ``small" and ``large" AdS black holes can be near-extremal. Equally importantly, in both cases the metric parameter a is by no means small, meaning that near-extremal AdS-Kerr black holes, whether large or small, are indeed significantly deformed away from the AdS-Schwarzschild geometry.

Now that we understand how to define the parameter ranges in which we are most interested, we can begin to explore the stability of these objects, in the sense discussed by Seiberg and Witten \cite{kn:seiberg}.

\addtocounter{section}{1}
\section* {\large{\textsf{3. Stability of the Boundary Theory}}}
Before turning to the [much more complicated] case of the AdS-Kerr-Newman geometry, let us consider why AdS itself is stable against the Seiberg-Witten effect.

The Euclidean version of the metric can be deduced directly from the global version of the Lorentzian AdS metric:
\begin{equation}\label{E}
\m{g(EAdS)\; =\; cosh^2\big(r/L\big)dt^2\;+\; dr^2\;+\;L^2\,sinh^2\big(r/L\big)\,\Big[d\theta^2 \;+\; sin^2(\theta)\,d\phi^2\Big]},
\end{equation}
where L as usual is the radius of curvature; this is of course just an unusual representation of [part of] hyperbolic space\footnote{Throughout this section, all metrics discussed are Euclidean.}. This part of hyperbolic space is however the physically relevant region for our purposes: it has a similar structure to that of a [Euclidean] topologically spherical AdS black hole. For that reason, we require t to be a periodic coordinate with period P [though of course that is not compulsory in this case].

Now the Euclidean action\footnote{In this work, ``Euclidean action" means the real function which has to be multiplied by $\sqrt{- 1}$ prior to being substituted into the path integral. It should be positive if the path integral is to converge.} of a BPS 2-brane wrapping a surface of the form r = constant takes a form which, up to an overall factor involving the tension, is purely geometric: it is proportional to the difference between the area of the brane and its [suitably normalized] volume; see \cite{kn:wittenyau}\cite{kn:maoz}\cite{kn:porrati}. The point made by Seiberg and Witten is that, in an asymptotically hyperbolic geometry, the competition between area and volume becomes very close as the radial coordinate increases: the leading terms cancel, and it is not clear a priori which subleading term will ``win". If, after analytic continuation, the Lorentzian action becomes negative in some region, then [the action being zero for a brane of minimal size] sufficiently large branes can minimize their action by moving into that region, and so branes produced by pair-production will not collapse under their own tension; the result will be a pair-production instability [of varying degrees of seriousness, depending on the extent of the region in question].

In the case of pure AdS, the area always exceeds the volume term: the Euclidean action of a BPS brane, with tension $\Theta$, located at a given value of r is
\begin{equation}\label{eq:F}
\m{E\$[AdS] \;=\; \Theta \Big[Area \;-\;{3\over L}Volume \Big]\;=\;4\pi P\Theta L^2e^{- r/L}sinh^2(r/L)}.
\end{equation}
The continuation to Lorentzian signature is trivial. One sees that the action is positive \emph{everywhere} for r $>$ 0, vanishing only at the origin. Thus AdS itself is completely stable against brane pair-production: any branes produced by pair-production will tend to collapse to the minimal size [zero in this case]. A similar result holds for the AdS-Schwarzschild geometry. In other cases, however, the action is negative for some values of r. As has been discussed in detail in \cite{kn:AdSRN}\cite{kn:triple}\cite{kn:73}, this is precisely what happens when AdS-Reissner-Nordstr$\ddot{\m{o}}$m black holes with \emph{flat} event horizons have a sufficiently large [but sub-extremal] charge: the action is actually negative and decreasing over an infinite domain, so a sufficiently large brane will grow indefinitely in an attempt to minimize the action. The system is completely unstable.

Seiberg and Witten observed that the AdS/CFT correspondence can be used to determine whether this happens at \emph{large} distances from the black hole. The idea is simply that any instability in that region of the bulk should be reflected in some kind of instability in the dual theory on the conformal boundary. Now for any conformally coupled scalar $\varphi$ in the boundary theory, the coupling term will be proportional to R$\varphi^2$, where R is the boundary scalar curvature, so negative scalar curvature at infinity tends to induce a tachyonic instability, at least for the zero mode. Conversely, positive R tends to stabilize the boundary physics. [No general statement is possible in the R = 0 case.] Seiberg and Witten argue that this is the dual version of the bulk instability discussed above. Thus the question of the sign of the action at large distances can be settled by duality.

It is easy to see how this works for AdS itself. The metric in (\ref{E}) induces on the surfaces r = constant the metric
\begin{equation}\label{FF}
\m{g(EAdS)(r  = constant)\; =\; L^2cosh^2\big(r/L\big)\Bigg[dt^2\;+\;tanh^2\big(r/L\big)\,\Big[d\theta^2 \;+\; sin^2(\theta)\,d\phi^2\Big]\Bigg]},
\end{equation}
which is obviously conformal to a metric with scalar curvature 2coth$^2$(r/L). This simply tends to a constant value, 2, as r tends to infinity; that is, the conformal structure at infinity [which, with the compactification of t, is now itself compact with topology S$^1 \times \, $S$^2$] is represented by a metric of positive scalar curvature. It follows by duality that there is no brane-induced instability in the large-r region of the AdS bulk, confirming our earlier conclusion for this region.

We wish to examine all of these questions in the Kerr-Newman case. For each fixed triple of parameters (a, M, Q), the AdS-Kerr-Newman metric defines a Euclidean counterpart [or ``gravitational instanton", see \cite{kn:hawrot}] of the form
\begin{equation}\label{C}
\m{g(EAdSKN) = {\Delta^E_r \over \rho_E^2}\Bigg[\,dt \; + \; {a \over \Xi^E}sin^2\theta \,d\phi\Bigg]^2\;+\;{\rho_E^2 \over \Delta^E_r}dr^2\;+\;{\rho_E^2 \over \Delta^E_{\theta}}d\theta^2 \;+\;{sin^2\theta \,\Delta^E_{\theta} \over \rho_E^2}\Bigg[a\,dt \; - \;{r^2\,-\,a^2 \over \Xi^E}\,d\phi\Bigg]^2},
\end{equation}
where
\begin{eqnarray}\label{eq:D}
\rho_{\m{E}}^2& = & \m{r^2\;-\;a^2cos^2\theta} \nonumber\\
\m{\Delta^E_r} & = & \m{(r^2-a^2)\Big(1 + {r^2\over L^2}\Big) - 2Mr - Q^2}\nonumber\\
\m{\Delta^E_{\theta}}& = & \m{1 + {a^2\over L^2} \, cos^2\theta} \nonumber\\
\m{\Xi^E} & = & \m{1 + {a^2\over L^2}.}
\end{eqnarray}
When (\ref{BBB}) is satisfied, the polynomial $\m{\Delta^E_r}$ has a positive root r$_0$, which of course is not equal to r$_{\m{eh}}$ in general, in fact it is larger. The range of r is now constrained by r $\geq$ r$_0$; one shows easily that r$_0$ is always larger than a, so the first member of (\ref{eq:D}) is always well-behaved.

The Euclidean geometry is completely regular provided that the cylinder spanned by t and $\phi$ is periodically identified under the action of the map
\begin{equation}\label{CROWN}
\m{(t,\;\phi)\;\rightarrow \;(t\,+\,P,\;\phi\,+\,\Phi )},
\end{equation}
where P and $\Phi$ are constants. These are not universal: they depend on a, M, and Q [see \cite{kn:hawrot}\cite{kn:solod}]. They are not independent: we have
\begin{equation}\label{CLOD}
\m{\Phi \;=\;{aP\Xi^E\over r_0^2 - a^2}}.
\end{equation}
As is explained in \cite{kn:hawrot}, P has a Lorentzian interpretation as the reciprocal of the Hawking temperature, T; equation (\ref{CLOD}) allows a similar Lorentzian interpretation\footnote{Under the analytic continuation [from Euclidean to Lorentzian signature], P is replaced by $-\,$i/T, and a by ia, so $\Phi$ remains real.} of $\Phi$.

A purely r-dependent conformal transformation shows that the conformal structure at infinity for this instanton is represented by the metric
\begin{equation}\label{J}
\m{g(EAdSKN)_{\infty}\;=\;d\tau^2 \;+\;{2\,(a/L)\,sin^2(\theta)\,d\tau d\phi \over 1 + (a/L)^2} \;+\; {d\theta^2 \over 1 + (a/L)^2cos^2(\theta)} \;+\; {sin^2(\theta)d\phi^2 \over 1 + (a/L)^2} },
\end{equation}
where $\tau$ is the dimensionless Euclidean time coordinate $\tau$ = t/L. Because of the identification under (\ref{CROWN}), this is a metric on the compact manifold S$^1 \times \,$S$^2$, and indeed it is the canonical metric [of positive scalar curvature 2] on that manifold when the angular momentum is zero.

The scalar curvature of g(AdSKN)$_{\infty}$ is given by
\begin{equation}\label{K}
\m{R(g(EAdSKN_{\infty})) \;=\; 2\Big(1\;-\;{a^2 \over L^2}\Big)\;+\;10\,{a^2 \over L^2}\,cos^2(\theta).}
\end{equation}
Notice that, at the equator [$\theta$ = $\pi$/2], the scalar curvature is smaller than its value when a = 0; however, it must be positive\footnote{By using a further conformal transformation which depends on the non-radial coordinates, one can in fact reduce the scalar curvature to a positive \emph{constant}, by an argument given in \cite{kn:besse}, page 124; note also that the Cotton tensor \cite{kn:cotton} of this metric vanishes, so it is conformally flat.} even there, because of the condition (\ref{BB}). Thus, according to the Seiberg-Witten criterion, there is no danger of negative brane actions at large distances from a Kerr-Newman black hole.

While this is reassuring, it does \emph{not} suffice to consider only the asymptotic region: even if the action is positive far from the black hole, it might be negative over a finite domain. Maldacena and Maoz \cite{kn:maoz} gave examples where this actually does happen. As they point out, this is much less serious than having an infinite quantity of negative action; what it means is that the system can be expected to evolve to some ``nearby" state with brane actions which \emph{are} positive everywhere. In our application, however, even this limited form of instability is unwelcome, because long-lived black holes which are very close to extremality have actually been observed. In other words, we need to show that the system can be static, and this means that we need to show evidence that the action is positive \emph{throughout} the bulk spacetime. Let us therefore turn to the computation of the action function.

\addtocounter{section}{1}
\section* {\large{\textsf{4. The Brane Action}}}
As in the case of BPS branes propagating in AdS itself, discussed above, the brane action is to be computed from the volume and area of a region in the Euclidean version of the geometry. The volume of a brane wrapping a surface at a fixed value of r $\geq$ r$_0$ in the Euclidean AdS-Kerr-Newman geometry is obtained by evaluating the integral of the square root of the determinant of the metric given in equation (\ref{C}), beginning from r = r$_0$. [This is the non-singular ``origin" of the Euclidean version of the spacetime; the brane action has to vanish there. It corresponds to the location of the event horizon in the Lorentzian version. Thus we are confining our discussion to the exterior of the black hole.] The determinant is given by\footnote{The determinant is most easily evaluated using the following elementary identity for 2$\times$2 matrices: det$\Bigg[ \left(
                       \begin{array}{cc}
                         \alpha_1^2& \alpha_1\beta_1 \\
                         \alpha_1\beta_1 & \beta_1^2 \\
                       \end{array}
                     \right)+\left(
                       \begin{array}{cc}
                         \alpha_2^2& \alpha_2\beta_2 \\
                         \alpha_2\beta_2 & \beta_2^2 \\
                       \end{array}
                     \right)\Bigg]=[\alpha_1\beta_2-\alpha_2\beta_1]^2$. This is to be applied to the two non-diagonal blocks corresponding to t and $\phi$.}
\begin{equation}\label{eq:G}
\m{\det\Big[g(EAdSKN)\Big] \;=\;{\rho_E^4 sin^2\theta \over (\Xi^E)^2}.}
\end{equation}
The area element is obtained by multiplying this determinant by $\m{\Delta^E_r/\rho_E^2}$ and taking the square root. The integrals in the $\phi$ and t directions are trivial\footnote{Note however that because $\Phi$ is proportional to a [equation (\ref{CLOD})], the result does not reduce to the action in the Schwarzschild case when a is reduced to zero; that case has to be treated separately \cite{kn:conspiracy}.} [see (\ref{CROWN})], and so we obtain a Euclidean brane action
\begin{equation}\label{eq:H}
\m{E\$(AdSKN) \;=\; P\Phi\Theta\Bigg[{\sqrt{\Delta^E_r} \over \Xi^E}\int_0^{\pi}\rho_E \,sin\theta \,d\theta\;-\;{3\over L\Xi^E}\int^r_{r_0}\int_0^{\pi}\rho_E^2 sin\theta \,d\theta \,dr \Bigg].}
\end{equation}
The integrals can be performed exactly, and the result, after analytic continuation of the parameters, is
\begin{eqnarray}\label{I}
\m{\$(AdSKN)} & \;=\; & \m{{a\Theta\over T^2(r_{eh}^2 + a^2)}\Bigg\{r\sqrt{(r^2+a^2)\Big(1 + {r^2\over L^2}\Big) - 2Mr + Q^2}\,\,\times \, \Bigg[\sqrt{1+{a^2\over r^2}}+ {r\over a}\, arcsinh{a\over r}\Bigg]}
\nonumber \\
& &
\;\;\;\;\;\;\;\;\;\;\;\;\;\;\;\;\;\;\;\;\;\;\;\;-\;\m{{2r^3\over L}\Bigg[1 + {a^2\over r^2}\Bigg]\;+\;{2r_{\m{eh}}^3\over L}\Bigg[1 + {a^2\over r_{\m{eh}}^2}\Bigg]\Bigg\},}
\end{eqnarray}
where T is the Hawking temperature of the black hole, $\Theta$ is the brane tension, and r$_{\m{eh}}$ is the radius of the event horizon.

\begin{figure}[!h]
\centering
\includegraphics[width=0.7\textwidth]{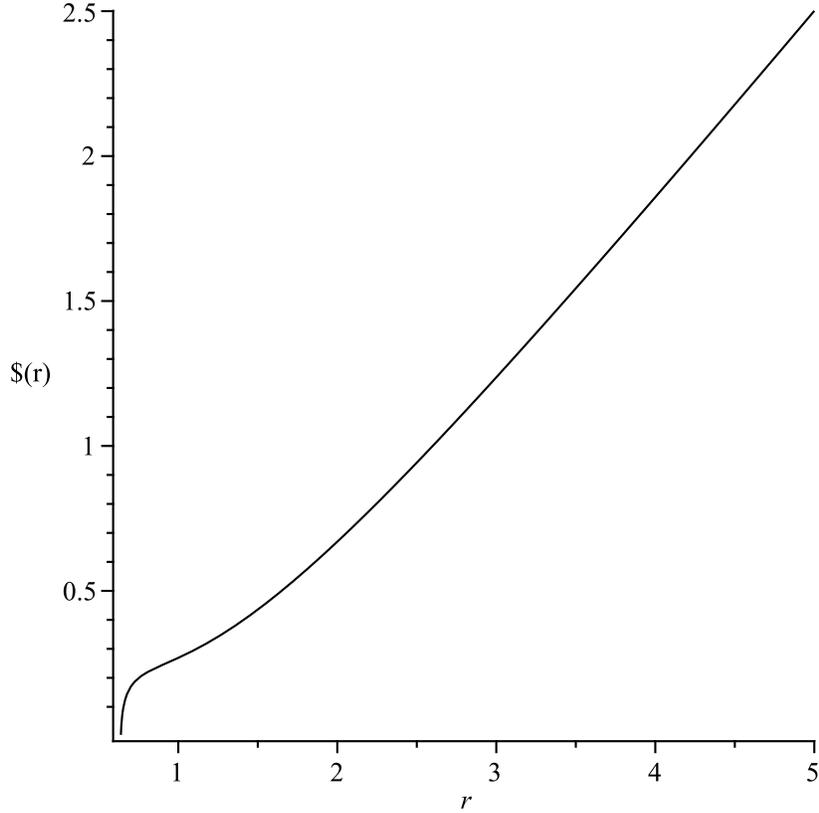}
\caption{[Scaled] brane action, Example S [$\m{M/L = 1, Q = 0, a/L = 1/\sqrt{2}.}]$ }
\end{figure}

We claim that this expression is positive ---$\,$ and in fact monotonically increasing as a function of r, so it has no minima except at r = r$_{\m{eh}}$ ---$\,$ everywhere outside the event horizon, for all \emph{physical} values of the parameters [meaning in particular that a/L $<$ 1]. We have not succeeded in finding an analytic proof of this statement [except for certain ranges of r, see below], but extensive numerical testing [see the Appendix] provides convincing evidence in support of it.
For the specific cases of Examples S and L discussed above, the graphs [of a convenient positive multiple of the action] are shown in Figures 2 and 3. These graphs agree with our claim, and are typical.

\begin{figure}[!h]
\centering
\includegraphics[width=0.7\textwidth]{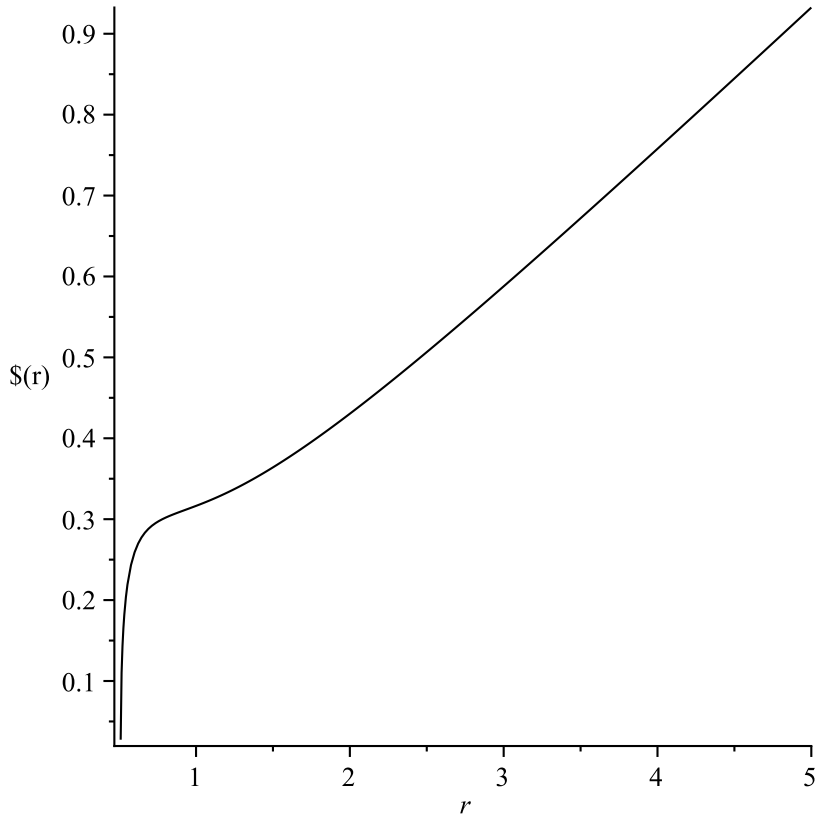}
\caption{[Scaled] brane action, Example L [$\m{M/L = 2, Q = 0, a/L = 0.98.}]$ }
\end{figure}

For large values of r, an analytic treatment is possible and illuminating. Inspecting the action function, one sees that the volume grows, to leading order, with a multiple of r$^3$; in fact, it grows slightly more rapidly, with a linear correction. However, the area \emph{also} grows slightly more rapidly than this same multiple of r$^3$ at large r, again with a linear correction. With the given coefficients, the two leading terms cancel exactly. But the coefficient of the linear term coming from the volume is exactly 2a$^2$/L, whereas the coefficient of the corresponding term coming from the area turns out to be different in general: we have
\begin{equation}\label{eq:HH}
\m{\$(AdSKN) \;=\;{a\Theta\over T^2(r_{eh}^2 + a^2)}\Bigg[\, rL \Bigg(1 - {2a^2 \over 3L^2}\Bigg)\;+\;{2r_{\m{eh}}^3\over L}\Bigg(1 + {a^2\over r_{\m{eh}}^2}\Bigg) - 2ML\;+\;O(1/r)\Bigg].}
\end{equation}
For all values of the angular momentum such that a$^2$/L$^2 <$ 1 [in fact, 3/2], the total coefficient of the term linear in r is strictly \emph{positive}. Thus the action will in fact grow linearly at large distances from the black hole; the slope decreases with increasing angular momentum, but it is always positive, due to the crucial constraint (\ref{BB}). Notice that this is consistent with Figures 2 and 3, where the graphs do rapidly become linear with positive slope. It is also consistent with our qualitative discussion, based on the Seiberg-Witten criterion, in the preceding section.

On the other hand, since the action function involves the square root of $\Delta_{\m{r}}$, which vanishes at the event horizon, its slope is infinite there, and, for values of r slightly larger than r$_{\m{eh}}$, the slope is large and positive, for all values of the parameters; again, this is reflected in the graphs in Figures 2 and 3. \emph{Thus, the action is certainly positive and increasing at both small and large r}. The claim, supported by strong numerical evidence, is that this is also true for intermediate values, so that there is no bulk instability.

In short: in no case is the brane action ever negative for any value of r, whether the black hole be small or large, no matter how close to extremality it may be: BPS branes in the AdS-Kerr-Newman spacetimes can never reduce their action by expanding. We conclude that the geometry of even the near-extremal AdS-Kerr-Newman black holes [with topologically spherical event horizons] is not sufficiently different from that of AdS itself to provoke the pair-production instability that does arise in other cases.

\addtocounter{section}{1}
\section* {\large{\textsf{5. Conclusion}}}
To summarize, then: the spacetime of an AdS-Kerr-Newman black hole, with topologically spherical event horizons, does indeed represent a significant distortion of the AdS-Schwarzschild geometry; at first sight, one would think that it is at least as deformed as the geometries considered by Murata et al. \cite{kn:murata}, which are unstable to the Seiberg-Witten effect. Surprisingly, however, the deformation in this case turns out to be harmless: angular momentum deforms the geometry, but not as much as it might have.

The way is therefore open for the programme proposed by Krishnan \cite{kn:krishnan}, to use string theory on large, near-extremal AdS-Kerr-Newman black holes in order to gain insight into the behaviour of observed black holes. While there are risks involved in using the gauge-gravity duality in this way to produce quantitative results, qualitative statements of this kind can still be valuable, as recently discussed in a different context in \cite{kn:mateos}\cite{kn:karch}. In that spirit, the principal qualitative result of this work may be stated as follows: contrary to expectations based on earlier studies of other kinds of black hole distortions, \emph{non-perturbative string effects do not destabilize very rapidly rotating black holes}.

We close with some remarks on ``small" AdS black holes. We have seen that, somewhat surprisingly, these objects are no more susceptible to Seiberg-Witten instability than their ``large" counterparts, though they tend to be unstable both to evaporation and to superradiance. Their description in terms of the AdS/CFT correspondence is obscure. Some exploratory work in this direction has however been done by Asplund and Berenstein \cite{kn:david}, who argue, in the context of the ``emergent quantum gravity" programme \cite{kn:beren}, that these holes are dual to a specific thermal but approximately isolated sub-sector of the field theory. In that work it is important that, even though the black hole is unstable, it should survive for a long time relative to its energy ---$\,$ it should be ``approximately metastable". Thus one needs to be able to show that there are no \emph{other} forms of instability which might shorten the lifetime to an unacceptable degree. Our work, in which we find no evidence for any further instability, is therefore compatible with these ideas.

\addtocounter{section}{1}
\section*{\large{\textsf{Acknowledgement}}}
The author is very grateful to Professor Bunster and the staff of the CECs, where this work was completed, for providing an environment conducive to research, for the opportunity to visit Chile, and for making him feel welcome.
\addtocounter{section}{1}
\section* {\large{\textsf{Appendix: Numerical Analysis of the Brane Action}}}
We claim that the action given in equation (\ref{I}) is positive everywhere outside the event horizon of the black hole. Since it is clearly zero at the event horizon, this will be proved if it can be shown that the action function is a strictly increasing function of r [as in Figures 2 and 3]. We proved that this is so for small and large values of r, but for intermediate values a numerical analysis is necessary.

We can simplify this task very considerably by defining a ``stripped down" action, SD$\$$(AdSKN), which is defined simply by removing all multiplicative and additive constants: that is,
\begin{eqnarray}\label{IRIS}
\m{SD\$(AdSKN)} & \;=\; & \m{r\sqrt{(r^2+a^2)\Big(1 + {r^2\over L^2}\Big) - 2Mr + Q^2}\,\,\times \, \Bigg[\sqrt{1+{a^2\over r^2}}+ {r\over a}\, arcsinh{a\over r}\Bigg]}
\nonumber \\
& &
\;\;\;\;\;\;\;\;\;\;\;\;\;\;\;\;\;\;\;\;\;\;\;\;-\;\m{{2r^3\over L}\Bigg[1 + {a^2\over r^2}\Bigg].}
\end{eqnarray}
Obviously we only need to show that this function is increasing [though of course it need not be positive]. We propose to do this by drawing three-dimensional graphs for a suitable range of parameters. We fix L = 1, and then draw graphs of the stripped down action, as a function of r and a, for fixed values of M. [The inequalities (\ref{BBB}) [relevant for small values of M] and (\ref{BB}) [relevant for larger M] discussed above allow us to determine the range of physical values for the angular momentum parameter a.] That is, for each M we can simultaneously plot SD$\$$(AdSKN) for the full range of possible values of a, and then inspect the graph to see whether it is increasing.

We have done this for all values of M between 0 and 10, increasing in steps of 0.25. It is immediately apparent that for each value of M and a, SD$\$$(AdSKN) [corresponding to the unlabelled axis in the graphs below] is indeed increasing as a function of r, and in fact it increases more and more steeply for larger values of M. A sample of these graphs, for M ranging from 1/4 to 4, is given in Figures 4 through 8. The graphs strongly support our claim.

\begin{figure}[!h]
\centering
\includegraphics[width=0.8\textwidth]{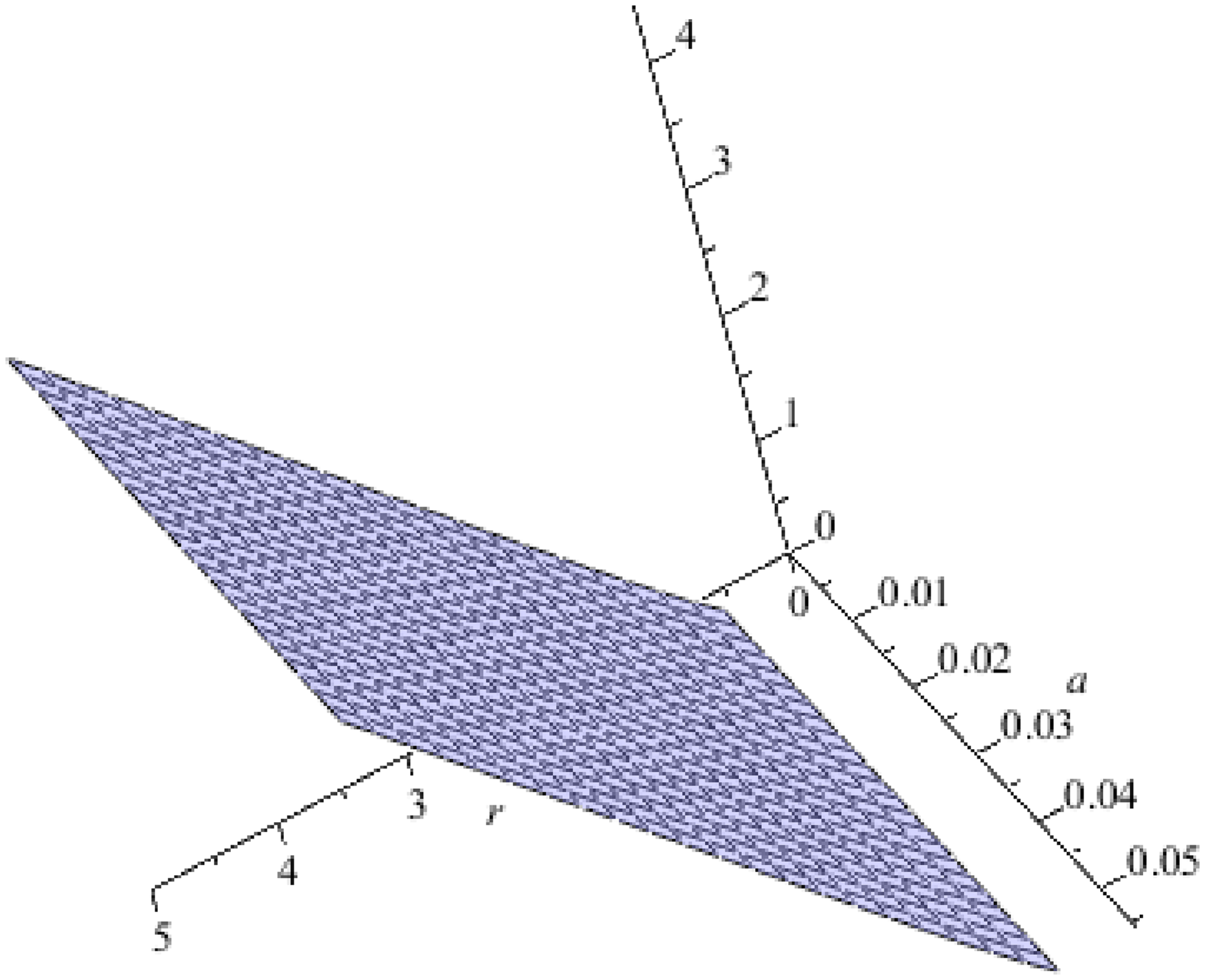}
\caption{SD$\$$(AdSKN), M/L = 1/4, all Physical Angular Momenta.}
\end{figure}
\begin{figure}[!h]
\centering
\includegraphics[width=0.8\textwidth]{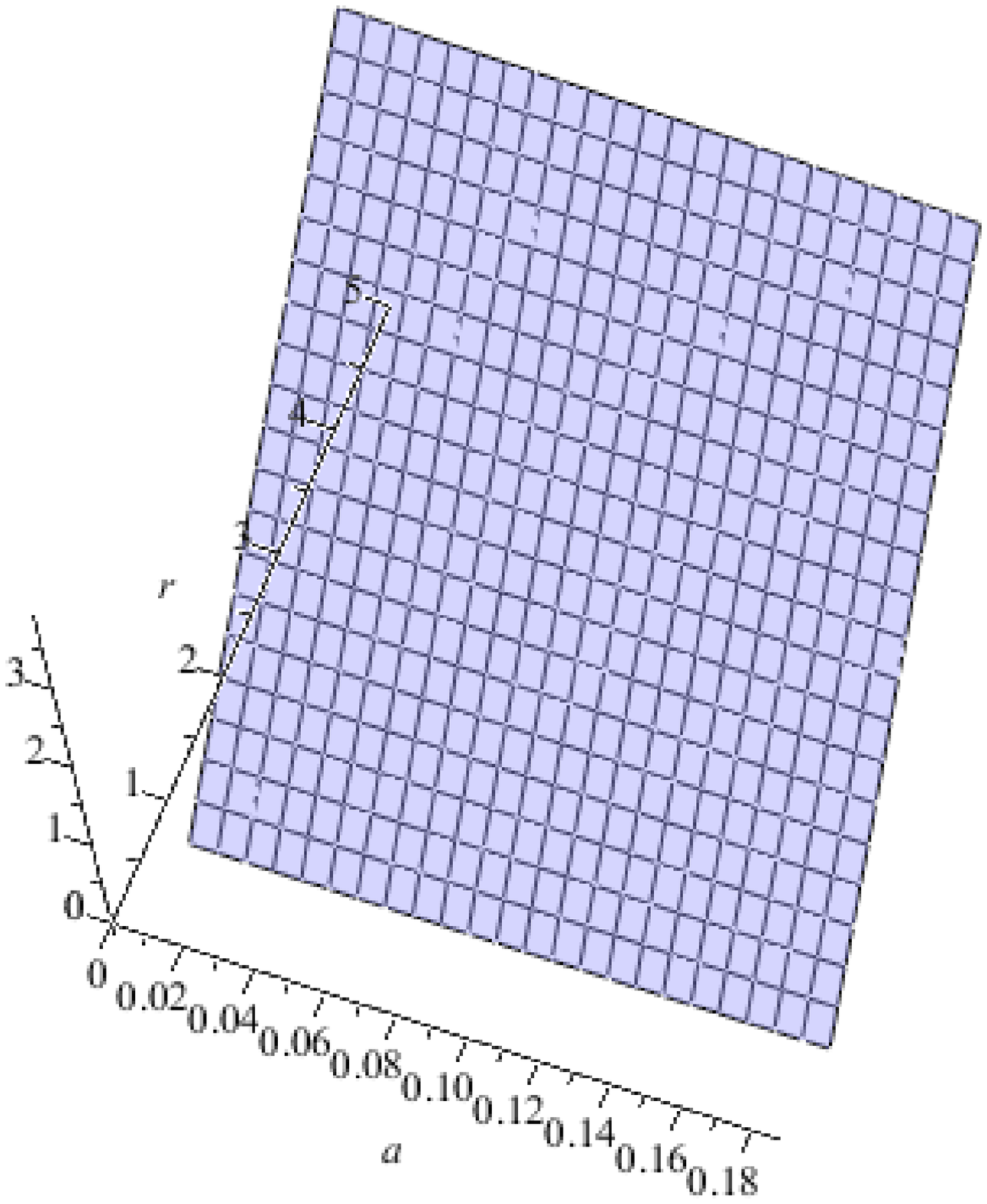}
\caption{SD$\$$(AdSKN), M/L = 1/2, all Physical Angular Momenta.}
\end{figure}
\begin{figure}[!h]
\centering
\includegraphics[width=0.8\textwidth]{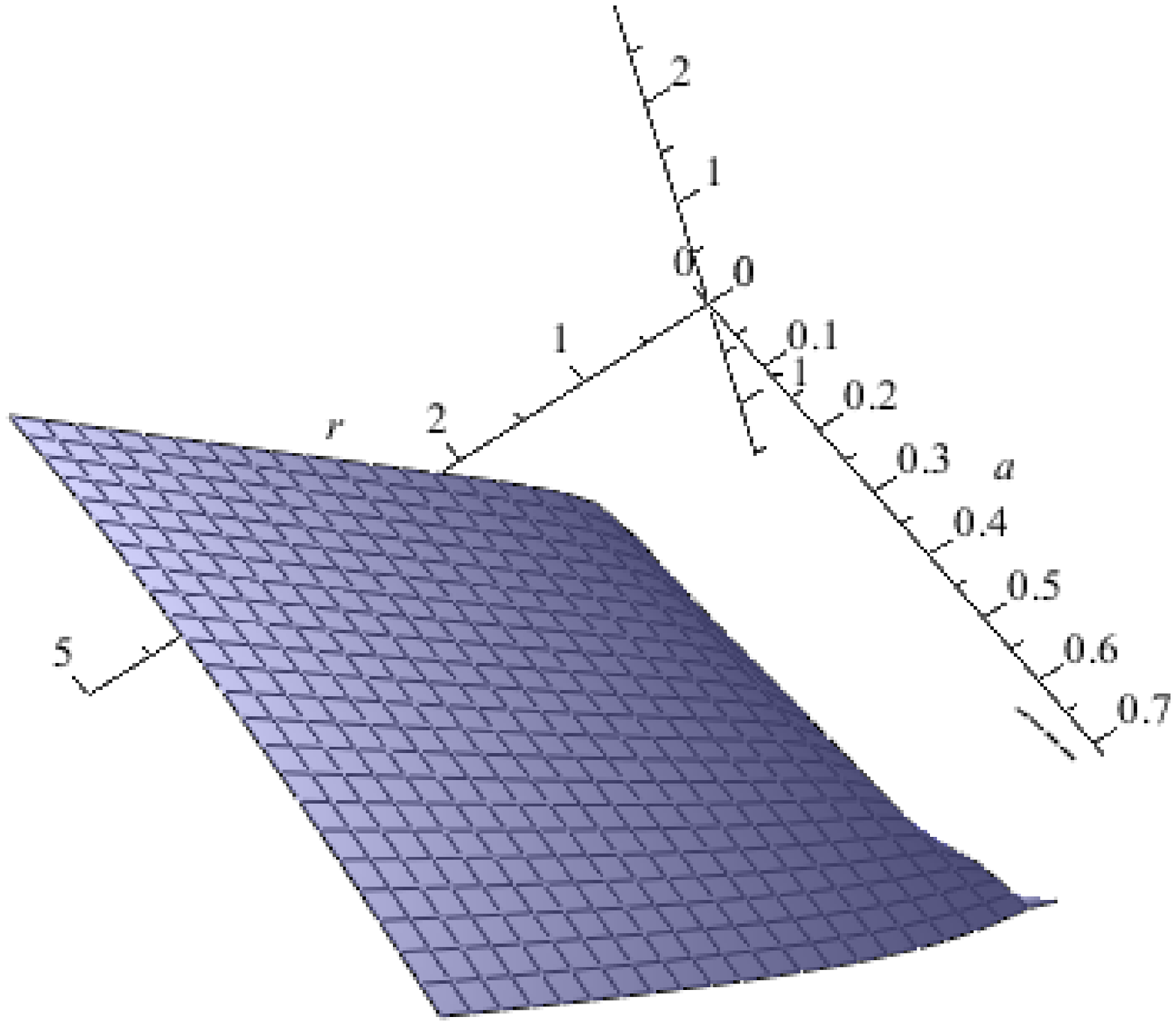}
\caption{SD$\$$(AdSKN), M/L = 1, all Physical Angular Momenta.}
\end{figure}
\begin{figure}[!h]
\centering
\includegraphics[width=0.8\textwidth]{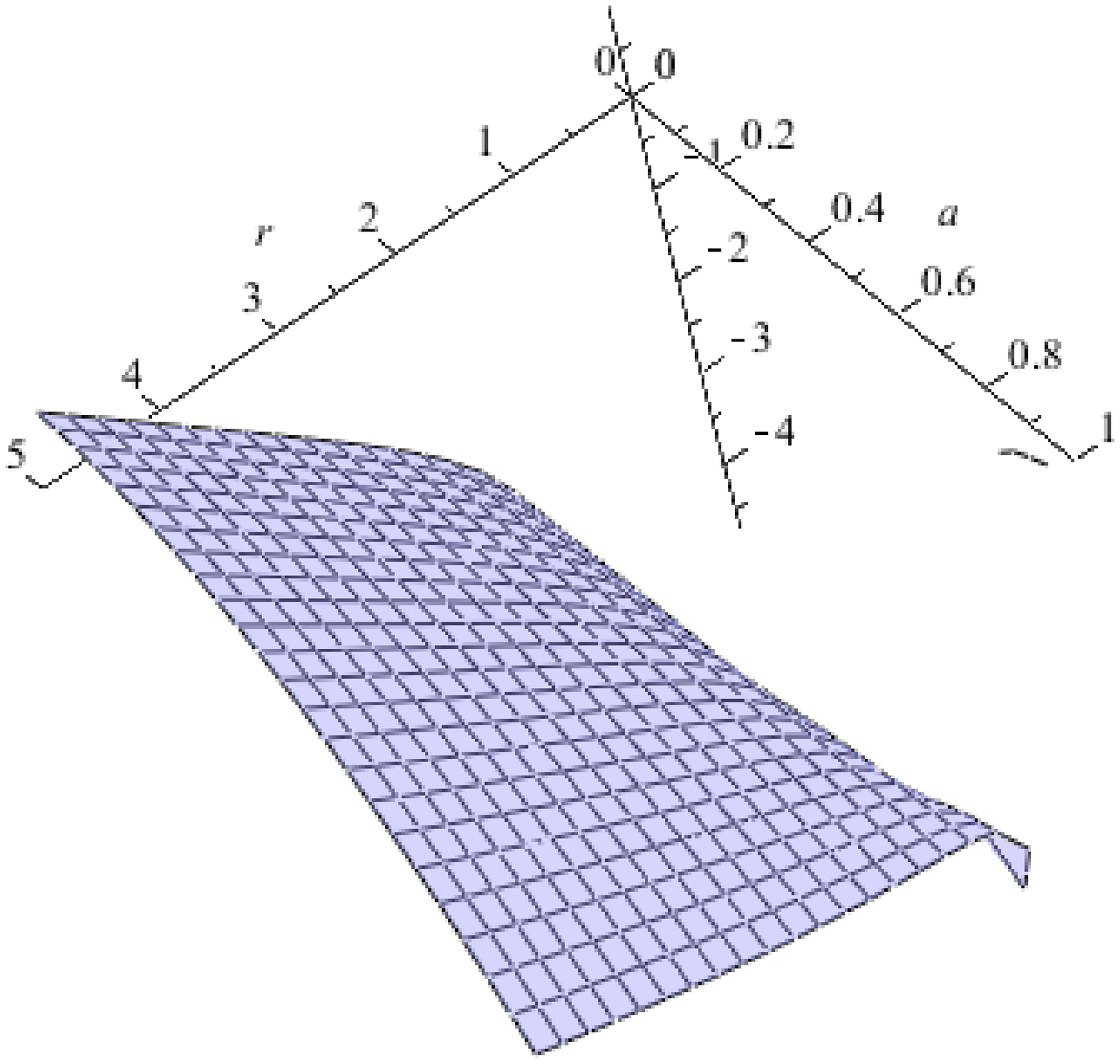}
\caption{SD$\$$(AdSKN), M/L = 2, all Physical Angular Momenta.}
\end{figure}
\begin{figure}[!h]
\centering
\includegraphics[width=0.8\textwidth]{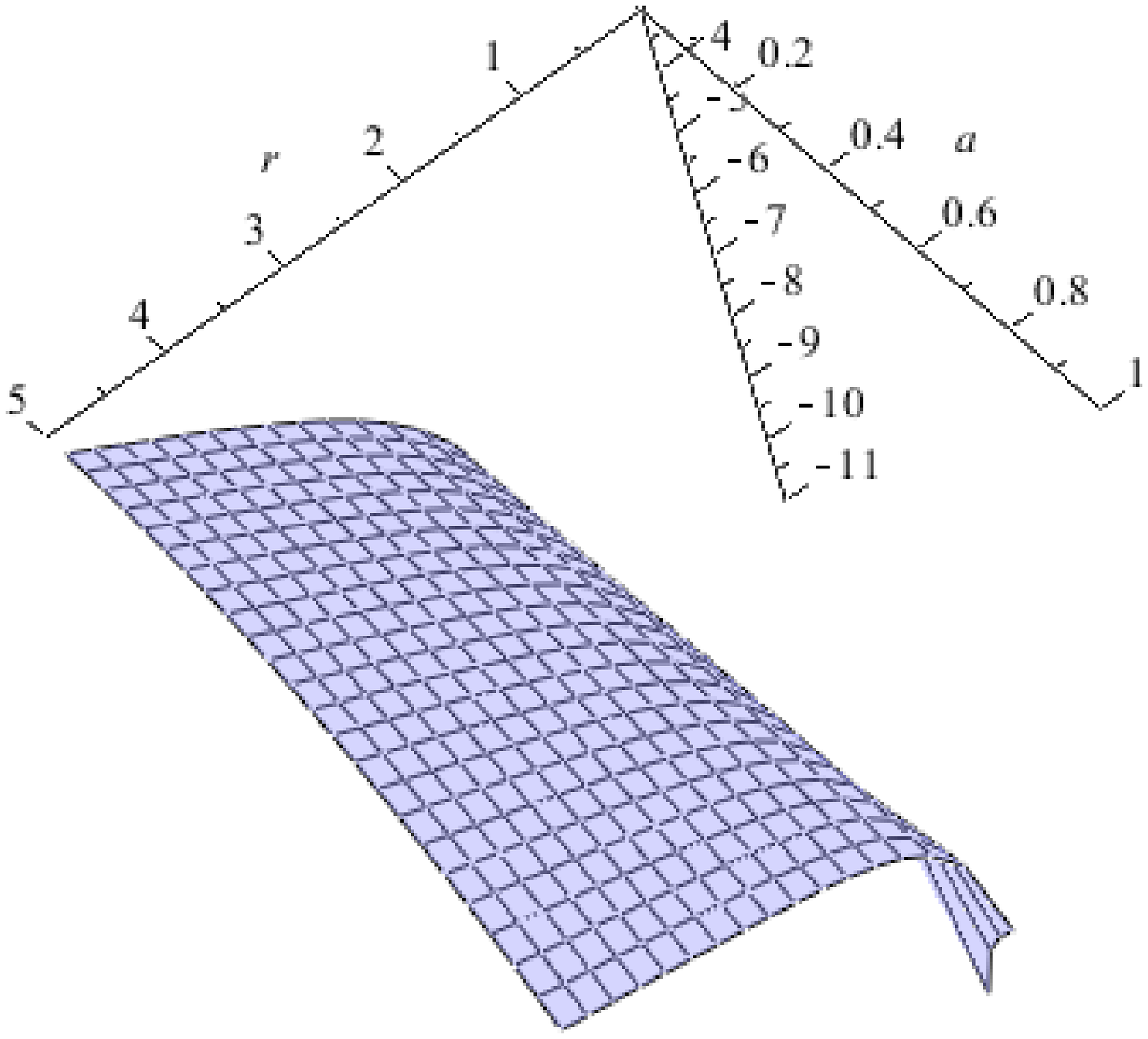}
\caption{SD$\$$(AdSKN), M/L = 4, all Physical Angular Momenta.}
\end{figure}
\end{document}